\documentclass{JHEP3}
\usepackage{graphics}
\usepackage{amssymb,amsfonts,epsfig,euscript,array,cite}

\newcounter{multieqs}

\newcommand{\be}{\begin{equation}}
\newcommand{\ee}{\end{equation}}
\newcommand{\eq}[1]{(\ref{#1})}

\newcommand{\bm}[1]{\mbox{\boldmath $#1$}}

\def\bd{\begin{document}}
\def\ed{\end{document}}
\def\nn{\nonumber}
\def\bea{\begin{eqnarray}}
\def\eea{\end{eqnarray}}
\let\bm=\bibitem
\let\la=\label

\def\npb#1#2#3{Nucl. Phys. {\bf{B#1}} #3 (#2)}
\def\plb#1#2#3{Phys. Lett. {\bf{#1B}} #3 (#2)}
\def\prl#1#2#3{Phys. Rev. Lett. {\bf{#1}} #3 (#2)}
\def\prd#1#2#3{Phys. Rev. {D \bf{#1}} #3 (#2)}
\def\cmp#1#2#3{Comm. Math. Phys. {\bf{#1}} #3 (#2)}
\def\cqg#1#2#3{Class. Quantum Grav. {\bf{#1}} #3 (#2)}
\def\nppsa#1#2#3{Nucl. Phys. B (Proc. Suppl.) {\bf{#1A}}#3 (#2)}
\def\ap#1#2#3{Ann. of Phys. {\bf{#1}} #3 (#2)}
\def\ijmp#1#2#3{Int. J. Mod. Phys. {\bf{A#1}} #3 (#2)}
\def\rmp#1#2#3{Rev. Mod. Phys. {\bf{#1}} #3 (#2)}
\def\mpla#1#2#3{Mod. Phys. Lett. {\bf A#1} #3 (#2)}
\def\jhep#1#2#3{J. High Energy Phys. {\bf #1} #3 (#2)}
\def\atmp#1#2#3{Adv. Theor. Math. Phys. {\bf #1} #3 (#2)}

\newcommand{\EQ}[1]{\begin{equation} #1 \end{equation}}
\newcommand{\AL}[1]{\begin{subequations}\begin{align} #1 \end{align}
\end{subequations}}
\newcommand{\SP}[1]{\begin{equation}\begin{split} #1 \end{split}\end{equation}}
\newcommand{\ALAT}[2]{\begin{subequations}\begin{alignat}{#1} #2
\end{alignat}\end{subequations}}
\def\beqa{\begin{eqnarray}}
\def\eeqa{\end{eqnarray}}
\def\beq{\begin{equation}}
\def\eeq{\end{equation}}

\def\N{{\cal N}}
\def\sst{\scriptscriptstyle}
\def\thetabar{\bar\theta}
\def\Tr{{\rm Tr}}
\def\one{\mbox{1 \kern-.59em {\rm l}}}

\def\a{\alpha}      \def\da{{\dot\alpha}}
\def\b{\beta}       \def\db{{\dot\beta}}
\def\c{\gamma}  \def\C{\Gamma}  \def\cdt{\dot\gamma}
\def\d{\delta}  \def\D{\Delta}  \def\ddt{\dot\delta}
\def\e{\epsilon}        \def\vare{\varepsilon}
\def\f{\phi}    \def\F{\Phi}    \def\vvf{\f}
\def\h{\eta}
\def\k{\kappa}
\def\l{\lambda} \def\L{\Lambda}
\def\m{\mu} \def\n{\nu}
\def\o{\omega}
\def\p{\pi} \def\P{\Pi}
\def\r{\rho}
\def\s{\sigma}  \def\S{\Sigma}
\def\t{\tau}
\def\th{\theta} \def\Th{\Theta} \def\vth{\vartheta}
\def\X{\Xeta}
\def\z{\zeta}

\def\cA{{\cal A}} \def\cB{{\cal B}} \def\cC{{\cal C}}
\def\cD{{\cal D}} \def\cE{{\cal E}} \def\cF{{\cal F}}
\def\cG{{\cal G}} \def\cH{{\cal H}} \def\cI{{\cal I}}
\def\cJ{{\cal J}} \def\cK{{\cal K}} \def\cL{{\cal L}}
\def\cM{{\cal M}} \def\cN{{\cal N}} \def\cO{{\cal O}}
\def\cP{{\cal P}} \def\cQ{{\cal Q}} \def\cR{{\cal R}}
\def\cS{{\cal S}} \def\cT{{\cal T}} \def\cU{{\cal U}}
\def\cV{{\cal V}} \def\cW{{\cal W}} \def\cX{{\cal X}}
\def\cY{{\cal Y}} \def\cZ{{\cal Z}}

\def\ua{\underline{\alpha}}
\def\ub{\underline{\phantom{\alpha}}\!\!\!\beta}
\def\uc{\underline{\phantom{\alpha}}\!\!\!\gamma}
\def\um{\underline{\mu}}
\def\ud{\underline\delta}
\def\ue{\underline\epsilon}
\def\una{\underline a}\def\unA{\underline A}
\def\unb{\underline b}\def\unB{\underline B}
\def\unc{\underline c}\def\unC{\underline C}
\def\und{\underline d}\def\unD{\underline D}
\def\une{\underline e}\def\unE{\underline E}
\def\unf{\underline{\phantom{e}}\!\!\!\! f}\def\unF{\underline F}
\def\unm{\underline m}\def\unM{\underline M}
\def\unn{\underline n}\def\unN{\underline N}
\def\unp{\underline{\phantom{a}}\!\!\! p}\def\unP{\underline P}
\def\unq{\underline{\phantom{a}}\!\!\! q}
\def\unQ{\underline{\phantom{A}}\!\!\!\! Q}
\def\unH{\underline{H}}

\def\As {{A \hspace{-6.4pt} \slash}\;}
\def\bs {{b \hspace{-6.4pt} \slash}\;}
\def\Ds {{D \hspace{-6.4pt} \slash}\;}
\def\ds {{\del \hspace{-6.4pt} \slash}\;}
\def\ss {{\s \hspace{-6.4pt} \slash}\;}
\def\ks {{ k \hspace{-6.4pt} \slash}\;}
\def\ps {{p \hspace{-6.4pt} \slash}\;}
\def\pas {{{p_1} \hspace{-6.4pt} \slash}\;}
\def\pbs {{{p_2} \hspace{-6.4pt} \slash}\;}

\def\Fh{\hat{F}}
\def\Vh{\hat{V}}
\def\Xh{\hat{X}}
\def\ah{\hat{a}}
\def\xh{\hat{x}}
\def\yh{\hat{y}}
\def\ph{\hat{p}}
\def\xih{\hat{\xi}}

\def\psit{\tilde{\psi}}
\def\Psit{\tilde{\Psi}}
\def\tht{\tilde{\th}}

\def\At{\tilde{A}}
\def\Qt{\tilde{Q}}
\def\Rt{\tilde{R}}
\def\Nt{\tilde{N}}

\def\at{\tilde{a}}
\def\st{\tilde{s}}
\def\ft{\tilde{f}}
\def\pt{\tilde{p}}
\def\qt{\tilde{q}}
\def\vt{\tilde{v}}
\def\nt{\tilde{n}}

\def\delb{\bar{\partial}}
\def\bz{\bar{z}}
\def\bD{\bar{D}}
\def\bB{\bar{B}}

\def\bk{{\bf k}}
\def\bl{{\bf l}}
\def\bp{{\bf p}}
\def\bq{{\bf q}}
\def\br{{\bf r}}
\def\bx{{\bf x}}
\def\by{{\bf y}}
\def\bR{{\bf R}}
\def\bV{{\bf V}}

\def\d{\delta}\def\D{\Delta}\def\ddt{\dot\delta}

\def\pa{\partial} \def\del{\partial}
\def\xx{\times}
\def\uno{\mbox{1 \kern-.59em {\rm l}}}

\def\trp{^{\top}}
\def\inv{^{-1}}
\def\dag{{^{\dagger}}}
\def\pr{^{\prime}}

\def\rar{\rightarrow}
\def\lar{\leftarrow}
\def\lrar{\leftrightarrow}

\newcommand{\0}{\,\!}      
\def\one{1\!\!1\,\,}
\def\im{\imath}
\def\jm{\jmath}

\newcommand{\tr}{\mbox{tr}}
\newcommand{\slsh}[1]{/ \!\!\!\! #1}

\def\vac{|0\rangle}
\def\lvac{\langle 0|}

\def\hlf{\frac{1}{2}}
\def\ove#1{\frac{1}{#1}}

\def\Box{\square}
\def\ZZ{\mathbb{Z}}
\def\CC#1{({\bf #1})}
\def\bcomment#1{}
\def\bfhat#1{{\bf \hat{#1}}}
\def\VEV#1{\left\langle #1\right\rangle}

\newcommand{\ex}[1]{{\rm e}^{#1}} \def\ii{{\rm i}}

\newcommand{\lrbrk}[1]{\left(#1\right)}
\newcommand{\sfrac}[2]{{\textstyle\frac{#1}{#2}}}

\font\mybb=msbm10 at 12pt
\def\bb#1{\hbox{\mybb#1}}

\font\myBB=msbm10 at 18pt
\def\BB#1{\hbox{\myBB#1}}

\title{Colliding Plane Waves in String Theory}

\author{Bin Chen\\
Interdisciplinary Center of Theoretical Studies, \\
Chinese Academy of Science, P.O. Box 2735\\
Beijing 100080, China\\
 \email{bchen@itp.ac.cn}}

\author{Chong-Sun Chu\\
Department of Physics,
 National Tsing-Hua University ,\\
 Hsin-Chu, 300 Taiwan\\and\\
Department of Mathematical Sciences, \\
University of Durham, Durham, DH1 3LE, UK\\
\email{Chong-Sun.Chu@durham.ac.uk}}

\author{Ko Furuta\\
Department of Physics, National Tsing-Hua University,\\
 Hsin-Chu 300, Taiwan\\
\email{furuta@phys.nthu.edu.tw} }

\author{Feng-Li Lin \\
Physics Division,
 National Center for Theoretical Sciences,
\\ National Tsing-Hua University,
 Hsin-Chu 300, Taiwan\\
\email{fllin@phys.cts.nthu.edu.tw}}

\date{\today}

\abstract{ We construct colliding plane wave solutions in higher
dimensional gravity theory with dilaton and higher form flux,
which appears naturally in the low energy theory of string theory.
Especially, the role of the junction condition in constructing the
solutions is emphasized. Our results not only include the previously known
CPW solutions, but also provide a wide class of new solutions that is not
known in the literature before. We find that late time curvature
singularity is always developed for the solutions we obtained in
this paper. This supports the generalized version of Tipler's
theorem in higher dimensional supergravity.}

\preprint{{\tt hep-th/0311135}}
\begin{document}


\section{Introduction}

The gravitational colliding plane wave (CPW) was first studied as
an exact solution of Einstein equation by Szekeres \cite{sz} and
Khan and Penrose  \cite{KP} in their pioneering papers and have
received much attention since then, see \cite{griffiths} and the
references therein. See also \cite{belinski} for an exposition on
the relation of CPW with Backlund transformation and inverse
scattering method.

Unlike the collision of waves in electromagnetic theory,
gravitational waves can interact nontrivially due to the nonlinear
nature of the Einstein equations. One of the intriguing feature of
CPW is the inevitable late time scalar curvature singularity which
signifies the non-linearity of the theory. The  curvature
singularities was initially discovered by Khan and Penrose in
their original paper \cite{KP}. It was then shown
\cite{Tipler,yu88} to be a general consequence of colliding plane
wave spacetime with plane symmetry. For a brief review of this
theorem, see, for example, \cite{rev}.  Similar null-like or
spacelike curvature singularities also arise in the big-bang
cosmology and in the black hole, which leads to the breakdown of
the classical gravity. It is hoped that further study of the CPW
may lead to a new understanding of this kind of the curvature
singularities.

CPW has appeared in various interesting physical settings. For
example it has been argued \cite{prebigbang,fkv,bv} that the
collision of gravitational plane waves could lead to initial
conditions of the primordial cosmological perturbations. In
\cite{Gowdy}, similar construction of CPW has lead to the new type
of inhomogeneous cosmology known as Gowdy universe. CPW has also
been employed as a useful approximation to study Planckian
scattering \cite{hooft}. The implication to the null-like
singularity of the Kerr black hole has   been discussed in
\cite{Ori}.

To obtain an exact solution of the Einstein equation which
describes the collision of gravitational plane waves, one first
divides the spacetime into 4 regions, namely, past P-region ($u<0, v<0$),
right R-region ($u>0, v<0$),
left L-region ($u<0, v>0$), and future F-region  ($u,v>0$)
which describes respectively, the
Minkowski space before the plane waves arrive, the incoming waves
from right and left, and the collision region. One then solves the
differential equation of motion within the interior of each
region. However, the existence of solutions to the differential
equations in each region is not guaranteed to describe the
collision of waves in the whole spacetime. As a physical solution,
it is necessary that these solutions can be joined to each other
in a ``physical'' way  at the (null) junctions. The physical
conditions can be translated into conditions on the metric and are
called the junction conditions.

\begin{figure}[ht]
\label{fig1}
\begin{center}
{\scalebox{1}{
\includegraphics{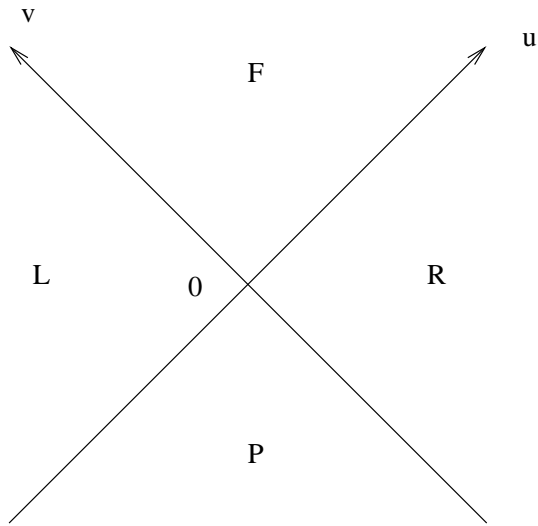}}
}
\end{center}
\caption{Spacetime in colliding plane wave problem. }
\end{figure}

  One of the generally accepted junction conditions in
general relativity is the Lichnerowicz condition which requires
that the metric is $C^1$ and piecewise $C^2$. This condition
guarantees that the curvature tensor is piecewise continuous
(P.C.) and nowhere suffers anything more than a `shock'
discontinuity, i.e. no delta function singularity. However, the
Lichnerowicz condition was found to be violated for the collision
of the  Khan-Penrose impulsive waves\cite{KP}. This was also realized
by Bell and Szekeres when they tried to construct the CPW with an
electromagnetic (EM) field shock wave profile \cite{BS}.
In the setting they considered, a EM potential is included. Therefore
a P.C. stress tensor is allowed; and so is
the same for the Ricci tensor. Bell and Szekeres thus proposed to consider
piecewise $C^1$ metric which admits a P.C.
Ricci tensor. Since for a general piecewise $C^1$ metric, the
Ricci tensor contains a $\d$-function singularity, it is necessary
to impose further conditions on the  piecewise $C^1$ metric such that the
$\d$-function singularity does not appear. These are precisely the O'Brien-Synge
(OS) junction conditions \cite{OSc}. See our appendix \ref{osj} for a
detailed exposition.

Apart from the above-mentioned $\d$-function singularity, it is
also possible that the curvature invariants $R$ and $R_2$
\footnote{ Here $R:=g^{\m\n} R_{\m\n}, R_2:= R^{\m\n}R_{\m\n},
R_4:=R^{\m\n\a\b}R_{\m\n\a\b}$ etc.. } blow up at the junction.
This should be avoided for a physical solution and generally
requires additional condition besides the
Lichnerowicz/O'Brien-Synge junction conditions. Our section 3
contains a general discussion on this point.

   Most of the discussions of CPW have been limited to the
4-dimensional gravity with or without a EM field. Dilatonic
gravitational CPW with a EM field has  been considered
\cite{Gurses1}. Higher dimensional generalization of the
Bell-Szekeres solution of Maxwell-Einstein gravity has
also been attempted recently \cite{higherBS}. As we will
demonstrate in section 4, the later solutions however violate the OS junction
conditions and are thus not
acceptable. Recently, Gutperle and Pioline \cite{GP} has tried to
construct the CPW in 10-dimensional IIB string theory with the
self-dual form flux; for latter convenience, we call the general
CPW with form flux the ``flux-CPW``. However, they found that the
curvature invariants blow up at the junction so that the solutions
cannot be used to describe the flux-CPW. The goal of this paper is
to construct regularly patched flux-CPW solution in string theory,
and we find that the key ingredient is to turn on the dilaton
field.

In this paper, we construct the flux-CPW solutions in higher
dimensional dilaton gravity, which includes the usual
10-dimensional II supergravity with either RR or NS form fluxes.
We find that by allowing a dilaton, the pole-like singularity at
the junctions in the curvature invariants can be avoided and we
obtained the higher dimensional flux-CPW solutions with dilaton.
Moreover, by adopting a new form of ansatz which is different from
the typical ones used by Bell-Szekeres, we obtain a new class of
flux-CPW solutions whose form has never been considered in the
literatures.

The organization of the paper is as follows. In section 2, we set
up and solve the differential equations in each of the four
regions. We find two type of solutions. One of which follows from
an ansatz whose form has never been considered in the literature.
In section 3, we analyze carefully the necessary (as well as some
additional, but uncompulsory) junction conditions that have to be
imposed on the metric. The discussions in this section are general
without referring to any particular solution. In section 4, we
impose these junction conditions on the solutions we find in
section 3 and  obtain physically acceptable flux-CPW solutions. We
also discuss the properties of these solutions. In particular, we
find that a curvature singularity is always developed in the
future. We conclude with a few discussions in section 5. Appendix
\ref{rr} contains some formulae for the Ricci and Riemann tensor
for the metric ansatz used for discussing CPW. Appendix \ref{osj}
contains a discussion of the $\d$-function singularity in
$R_{\m\n\a\b}, R_4, R_{\m\n}$ for a general piecewise $C^1$
metric. We show explicitly that the absence of $\d$-function
singularity in the Ricci tensor is precisely the OS junction
conditions. In appendix \ref{cpw-fluxless} we give solution for
the pure gravitation CPW and dilatonic CPW without flux. Appendix
\ref{bb} gives some details about the boundary behavior for the
CPW solutions we obtained. Finally in appendix \ref{brink}, we
give details on how the CPW solutions written in the Rosen
coordinates appear in the Brinkmann coordinate. In particular
we show that the singular structure of $R_4$ at the junction is
the same as the wavefront profile of the wave when written in the
Brinkmann coordinates.

\section{Equation of Motion and their Solution}

Consider the following action with the metric, dilaton and a
$(n+1)$-form field strength in $D$-spacetime dimensions
\be
S= \int d^D x \sqrt{-g} \biggl(R - g^{\m\n}\del_\m\phi \del_\n \phi -
\frac{1}{2(n+1)!}e^{a \phi} F^2\biggr),
\ee
where $F^2:= F_{\m_1 \cdots  \m_{n+1}} F^{\m_1 \cdots \m_{n+1}}$
and $a$ is the dilaton coupling constant. This action describes a
sector of the low energy effective action of string theory in the
Einstein frame. The values of $a$ depends on which string theory
we are working with and on the nature of the $(n+1)$-form field
strength.
The equation of motions are given by
\bea
&&R_{\m\n} = \del_\m\phi \del_\n \phi + \frac{1}{2 n!} e^{a \phi}
\biggl(F_{\m \m_1\cdots  \m_{n}} F_\n{}^{\m_1 \cdots \m_{n}} -
\frac{n}{(n+1)(n+m)}g_{\m\n}F^2\biggl),  \\
&&\del_\m(\sqrt{-g} e^{a \phi}F^{\m \m_1\cdots  \m_{n}}) =0, \\
&& \frac{1}{\sqrt{-g}}\del_\m(\sqrt{-g} g^{\m\n}\del_\n \phi) =
\frac{a}{4(n+1)!}  e^{a \phi} F^2. \eea

\subsection{CPW ansatz for the equation of motion}

Consider the following ansatz for the metric
\be \label{metric-ansatz}
ds^2 = 2 e^{-M} dudv + e^A \sum_{i=1}^n dx_i^2 +  e^B \sum_{j=1}^m
dy_j^2,
\ee
where $D=2+n+m$; and the nonzero components of the $(n+1)$-form
flux,
\be
F_{u x_1\cdots x_n} = C_u, \quad F_{v x_1\cdots x_n} = C_v.
\ee
We will take the functions $M, A, B, C$ as well as the dilaton
field $\phi$ to be function of $u,v$. The Einstein equations take
the form:
\bea
&& nA_{uu} +m B_{uu} + nM_u A_u + mM_u B_u + \frac{1}{2}(n A_u^2 +
m
B_u^2) = -2 \phi_u^2 -e^{a\phi -nA}C_u^2, \label{ruu}\\
&& nA_{vv} +m B_{vv} + nM_v A_v + mM_v B_v + \frac{1}{2}(n A_v^2 +
m
B_v^2) = -2 \phi_v^2 -e^{a\phi -nA}C_v^2, \label{rvv}\\
&&-M_{uv} + \frac{n}{2}A_{uv} + \frac{m}{2}B_{uv} +
\frac{1}{4}(nA_uA_v+mB_uB_v) = - \phi_u\phi_v +
\frac{1}{2}\frac{n-m}{n+m}  e^{a\phi -nA}C_u C_v, \;\;\;\;\;\; \label{ruv}\\
&&2A_{uv} +n A_u A_v +\frac{m}{2}(A_u B_v +A_v B_u) =
-\frac{2m}{n+m}e^{a\phi -nA}C_u C_v, \label{rxx}\\
&&2B_{uv} +m B_u B_v + \frac{n}{2}(A_u B_v +A_v B_u) =
\frac{2n}{n+m}e^{a\phi -nA}C_u C_v. \label{ryy}
\eea
These follow from the $R_{uu}, R_{vv}, R_{uv}, R_{xx},
R_{yy}$-equations (in this order). The equation of motion for
dilaton and $n$-form potential are given by
\bea
&& 2 C_{uv} +(a\phi - \frac{1}{2}(nA-mB))_uC_v +(a\phi -
\frac{1}{2}(nA-mB))_v C_u =0,\label{e1}\\
&& \phi_{uv} + \frac{1}{4}(nA+mB)_u \phi_v + \frac{1}{4}(nA+mB)_v
\phi_u = \frac{a}{4}e^{a\phi - nA} C_u C_v . \label{e2}
\eea
Here we have abbreviated the derivatives by a subscript, e.g. $A_u
= \del_u A$. Note that equation \eq{ruv} is a consequences of the
other equations and is not an independent equation. So in the
following we will not write it anymore.

It is convenient to define
\be\label{UV}
U = \frac{1}{2}(n A + m B), \quad V = \frac{1}{2}(n A - m B).
\ee
The equations \eq{ruu}, \eq{rvv}, \eq{rxx}, \eq{ryy} become
\bea
&& U_{uu}+M_u U_u + \frac{m+n}{4mn} (U_u^2+V_u^2) +
  \frac{m-n}{2mn}U_u V_u = - \phi_u^2 - \frac{1}{2} e^{a\phi
    -nA}C_u^2, \label{vuu}\\
&&U_{vv}+M_vU_v + \frac{m+n}{4mn} (U_v^2+V_v^2) +
  \frac{m-n}{2mn}U_v V_v = - \phi_v^2 - \frac{1}{2} e^{a\phi
    -nA}C_v^2, \label{vvv}\\
&& U_{uv} +U_u U_v =0, \label{U}\\
&& V_{uv} + \frac{1}{2}(U_u V_v+ U_v V_u) =
-\frac{mn}{m+n}e^{a\phi -nA}C_u
  C_v, \label{V}
\eea
Equation \eq{U} says that $e^U$ is a free field and the general
solution to it is
\be
U = \log(f(u)+g(v)),
\ee
where $f,g$ are arbitrary functions. We will choose them to be
monotonic functions. One may  treat $f,g$ as coordinates
alternative to $(u,v)$.

It is convenient to change from the pair of variables  $V, \phi$
to $E,X$ defined as  follows:
\be \label{EX}
E= V- a \phi, \quad X= \phi + \delta a V,
\ee
where the constant $\d$ is chosen to be
\be
\d : = \frac{n+m}{4nm} \leq\frac{1}{2}.
\ee
With this choice and some linear combinations, the equations
\eq{e1}, \eq{e2} and \eq{V} take the simple form in terms of the
$(f,g)$-coordinates
\bea
&& (f+g) X_{fg} + \frac{1}{2}( X_f + X_g) =0. \label{feqX}\\
&& 2 C_{fg} -E_f C_g -E_g C_f =0, \label{feqC}\\
&& (f+g) E_{fg} + \frac{1}{2}(E_f + E_g) = - \frac{\a }{4 \d}
e^{-E} C_f C_g\;, \label{feqE}
\eea
where
\be
\a := 1+ a^2 \d\;.
\ee

The equations \eq{vuu} and \eq{vvv} can be integrated to give $M$.
In terms of the $(f,g)$-coordinates, they can be written as
\bea
&& S_f + {1\over2} e^{-E} C_f^2 + {\d \over \a}(f+g) E^2_f + {1
\over  \a}(f+g) X_f^2= 0\;, \label{feqS1}
\\
&& S_g +  {1\over2} e^{-E} C_g^2 + {\d \over \a}(f+g) E^2_g + {1
\over  \a}(f+g) X_g^2= 0\;, \label{feqS2}
\eea
where we have defined
\be \label{SM}
S = M-(1-\d)\log(f+g) + \log(f_u g_v) + \eta V\;,
\ee
and
\be
\eta:={m-n\over 2mn}\;.
\ee
We remark that $X=0$ when there is no dilaton. In this case, $E =
V$, $\a=1$ and \eq{feqS1}, \eq{feqS2} reduce to the equation of
non-dilatonic gravity with form flux. Therefore the last term in
\eq{feqS1}, \eq{feqS2} can be identified as a contribution of the
dilaton field to $S$, i.e. to the metric component $e^{-M}$. As we
will show in section 4, this contribution plays an essential role
to allow for physically acceptable CPW solution, i.e. one that
satisfies the junction conditions as spelled out in section 3.

 The inverse relation of \eq{EX} is
\be \label{Vphi}
V= {1\over \a}(E+aX), \quad \phi =  {1\over \a}(X - a \d E).
\ee

Summarizing, the equation of motion for our system is given by the
equations \eq{feqX}-\eq{feqS2}. Our goal is to solve them for the
variables $(S,E,X,C)$. Then using \eq{Vphi} \eq{SM}, we can solve
for $(M, A, B, C, \phi)$. In the next subsection, we will give two
particular families of solution to the differential equations.

\subsection{Two flux-CPW solutions to  the equation of motion}

The case without form flux is easy to solve and the solutions are
given in the appendix \ref{cpw-fluxless}. In the following we will assume that
potential $C$ is nonzero. To solve for the most general solution
for the above set of coupled differential equations is very
difficult. In the following we will give two different solutions
using two different form of ansatz.

We first consider the $X$-equation \eq{feqX}. We note that it
takes the same form as  in the standard pure gravitational plane
wave collision \cite{sz,griffiths}, and it can be  solved by the
Khan-Penrose-Szekeres solution:
\be
X=\k_1 \log{w-p \over w+p} +\k_2 \log{r-q\over r+q}\;,
\label{solnX}
\ee
where $\k_1$ and $\k_2$ are integration constants and
\be
p:=\sqrt{\frac{1}{2}-f}\;, \quad q:=\sqrt{\frac{1}{2}-g}\;,\quad
r:=\sqrt{\frac{1}{2}+f}\;,  \quad w:=\sqrt{\frac{1}{2}+g}\;.
\ee

\bigskip

\underline{Solution I: $(pqrw)$-type}

Our solution I is given by the following ansatz for $E$ and $C$:
\be \label{solnE}
E=\log{rw+pq\over rw-pq}\;, \qquad C=\gamma (pw-rq),
\ee
which solves \eq{feqC} automatically and from \eq{feqE} we find
that
\be
\gamma^2={8\d \over \a}\;.
\ee
After integrating \eq{feqS1} and \eq{feqS2} with $X$ given by
\eq{solnX} to get
\bea \label{SolnSI}
S=&&b_1\log(1-2f)(1+2g)+b_2\log(1+2f)(1-2g)+(b_3-1+\d)\log(f+g) \nn\\
&&+{2\k_1\k_2\over \a}\log({1\over2}+2fg+2pqrw),
\eea
where
\be
\label{biI}
 b_1={\k_1^2+\d \over\a}\;, \qquad b_2={\k_2^2+\d \over \a}\;, \qquad
b_3=1-\d-{\d+(\k_1+\k_2)^2 \over \a}\;,
\ee
and using \eq{SM} we find that
\bea\label{solnM}
e^{-M}&=& f_u  g_v [(1-2f)(1+2g)]^{-b_1}[(1+2f)(1-2g)]^{-b_2}
(f+g)^{-b_3}
\nn\\
&\cdot & [{1\over2}+2 f g + 2 pqrw]^{-{2\k_1 \k_2\over \a}} \;
({rw+pq \over rw-pq})^{\eta \over \a} \; \Bigl[({w-p\over
w+p})^{\k_1} ({r-q\over r+q})^{\k_2}\Bigr]^{ {a\eta \over \a}}\;.
\eea
The other components of the metric are given by
\bea
&& e^{nA} = (f+g)({rw+pq \over rw-pq})^{1\over \a}\;
\Bigl[({w-p\over w+p})^{\k_1}({r-q\over r+q})^{\k_2}\Bigr]^{a
\over\a},
\label{solnA}\\
&& e^{mB} = (f+g)({rw+pq \over rw-pq})^{-{1\over \a}} \;
\Bigl[({w-p\over w+p})^{\k_1}({r-q\over r+q})^{\k_2}\Bigr]^{-{a
\over \a}}
\label{solnB}
\eea
and the dilaton field is given by
\be\label{solnd}
e^\phi = ({rw+pq \over rw-pq})^{-{\d a\over \a}}\;
\Bigl[({w-p\over w+p})^{\k_1}({r-q\over r+q})^{\k_2}\Bigr]^{1
\over \a}.
\ee

The above \eq{solnX}, \eq{solnE}-\eq{solnd} solve the equations of
motion for the F-region  and represents a two-parameters
family of solutions depending on the constants $\k_1$ and $\k_2$.
They still solve the EOM for the L-region, or for the
R-region, or for the P-region if one
do the following replacements:
\bea\label{fg}
&& f(u)= f_0, \quad f_u(1-2 f)^{-b_1}|_{f=f_0}=-1 \quad \mbox{for}
\quad u<0,\\
\label{fga} && g(v)= g_0, \quad g_v(1-2 g)^{-b_2}|_{g=g_0}=-1 \quad
\mbox{for} \quad v<0,
\eea
for some constants $f_0, g_0 $. In the next section, we will
discuss the patching of the solutions of the different regions.
This allow us to fix the values of
\be \label{fg0}
f_0 = g_0 = 1/2
\ee
and put constraints on the  parameters  $\k_1, \k_2$.

\bigskip

\underline{Solution II: $(f\pm g)$-type}

Next we present our solution II. It is obtained by taking the
following ansatz for $E$ and $C$ whose dependence in $f,g$ are of
the form $f\pm g$:
\be
E=E(f+g)\;, \qquad C=C(f-g)\;.
\ee
Equation \eq{feqC} then gives
\be \label{solnC2}
C=\c (f-g)\;,
\ee
for some constant $\c$, and \eq{feqE} can be solved by
\be \label{solnE2}
E = \log\Bigl[{\a \c^2 \over 8 \d c_1^2}(f+g) \cosh^2(c_1 \log{c_2
\over f+g})\Bigr],
\ee
where  $c_1$, $c_2$ are integration constants. Without loss of
generality one can take $c_1, c_2 >0$. One can then integrate
\eq{feqS1} and \eq{feqS2} to get
\bea
\label{SolnSII}
S&=&b_1\log(1-2f)(1+2g)+b_2\log(1+2f)(1-2g)+(b_3-1+\d +
\frac{\eta}{\a})\log(f+g)
\nn \\
&+&{2\k_1\k_2\over \a}\log({1\over2}+2fg+2pqrw)+{\d\over
\a}[(1+4c_1^2)\log c_2-4\log\cosh(c_1 \log{c_2\over f+g})]\;\;\;\;
\eea
and in this case,
\be
\label{biII}
b_1={\k_1^2\over \a},\quad b_2={\k_2^2\over \a}, \quad
b_3=1-\d-{\d(1+4c_1^2) +(\k_1+\k_2)^2\over \a}-{\eta \over \a}\;.
\ee

Finally we obtain
\bea \label{solnM2}
e^{-M}&=& a_0 f_u g_v [(1-2f)(1+2g)]^{-b_1} [(1+2f)(1-2g)]^{-b_2}
(f+g)^{-b_3} \cosh^{a_4}(c_1 \log{c_2 \over f+g})\nonumber\\
&\cdot& [{1\over2}+2fg+2pqrw]^{-{2\k_1\k_2\over \a}}
\Bigl[({w-p\over w+p})^{\k_1}({r-q\over r+q})^{\k_2}\Bigr]^{
{a\eta \over \a}},
\eea
where
\be
a_0 = c_2^{-{\d\over \a}(1+4c_1^2)} ({\a \c^2\over 8 \d
c_1^2})^{\eta \over \a}, \qquad a_4={4\d\over \a}+{2\eta \over
\a}\;.
\ee
The other components of the metric are given by
\bea
&& e^{nA} = ({\a \c^2\over 8 \d c_1^2})^{1\over \a }
(f+g)^{1+{1\over \a}} \cosh^{2\over\a}(c_1 \log{c_2 \over f+g})\;
\Bigl[({w-p\over w+p})^{\k_1}({r-q\over
r+q})^{\k_2}\Bigr]^{a\over\a},
\label{solnA2}\\
&&e^{mB} =({\a \c^2\over 8 \d c_1^2})^{-{1\over \a} }
(f+g)^{1-{1\over \a}} \cosh^{-{2\over\a}}(c_1 \log{c_2 \over
f+g})\; \Bigl[({w-p\over w+p})^{\k_1}({r-q\over
r+q})^{\k_2}\Bigr]^{-{a\over\a}} ,
\label{solnB2}\eea
and the dilaton field is given by
\be\label{solnd2}
e^\phi = ({\a \c^2\over 8 \d c_1^2})^{-{\d a \over \a }}
(f+g)^{-{\d a \over \a }} \cosh^{-{2 \d a \over \a}}(c_1 \log{c_2
\over f+g}) \; \Bigl[({w-p\over w+p})^{\k_1}({r-q\over
r+q})^{\k_2} \Bigr]^{1\over \a}.
\ee

The above \eq{solnX}, \eq{solnC2}-\eq{solnd2} give a
five-parameters family of solution in the F-region. As in the
$(pqrw)$-type solution given above, they still solve the
differential equations in the L-, R- and in the P-region if we
take \eq{fg}, \eq{fga} and \eq{fg0}. Patching of the solutions
together will put constraints on the parameters
$\gamma,c_1,c_2,\k_1,\k_2$.


\section{Junction Conditions of Metric}

In the last section, we have solved the equation of motion in the
different regions, and one need to paste together these solutions
across the boundaries. To be a physically acceptable solution, it
is necessary that the  metric $g_{\m\n}$ (spacetime) must be
continuous and invertible. What about the derivatives of the
metric? what about the $R_{\m\n}$, $R_{\m\n \a\b}$ and the various
curvature invariants $R, R_2, R_4$ etc.? What kind of
conditions shall we impose on them?
\footnote{We will
distinguish and analysis the behaviour of various quantities (e.g.
the first derivatives of the metric, $R_{\m\n}$, $R_{\m\n\a\b}$
etc.) at the junction. In general a physical quantity can take the
following form across a junction, say $u=0$,
\be
h(u) = h^{(0)}(u) + h^{(1)}(u)\Theta (u)+ h^{(2)}(u)\delta(u),
\ee
where $h^{(i)}(u)$ are continuous functions for $u>0$ and $u<0$,
and $ h^{(0)}(u)$ is continuous across $u=0$. The quantity $h$ is
continuous if there is no $\Th$ or $\d$-function, i.e. $
h^{(1)}(0) =h^{(2)}(0) =0$. $h$ is P.C.  if $h^{(2)}(0) =0$ and $
h^{(1)}(0) \neq 0$.  Note that we allow the jump $h^{(1)}(0)$ to
be both finite or infinite. When $h^{(2)}(0) \neq 0$, we have a
Dirac delta function singularity at the junction.}

It is natural to allow the Ricci tensor to be P.C.. This is particularly clear
in the presence of a form flux. In this case, the stress
tensor does not need to be continuous in spacetime and can
generally be P.C. with jump across the junctions. Let $\cS$ be a
hypersurface where the form flux $F_{[n+1]}$ is discontinuous across.
It can be shown that $\cS$ has to be a null surface. See the
appendix \ref{osj}. Through Einstein equation, this
implies that $R_{\m\n}$ can have jump across $\cS$ too.

It is also natural to assume that the energy momentum tensor does not suffer
anything more than a `shock' discontinuity, i.e. no delta function
singularity. Therefore a first requirement on $R_{\m\n}$ is that it should
contain no delta function singularity.

To achieve this, one may demand the metric to be $C^1$,
and at least piecewise $C^2$. This is the Lichnerowicz
condition. While this is sufficient, it is not necessary. As it turns
out, the absence of delta function singularity in $R_{\m\n}$ is also
possible for metric that is piecewise $C^1$ and that satisfy a certain
special condition. Generally, a piecewise $C^1$ metric
induces a Dirac delta
function singularity  for $R_{\m\n\a\b}$
and in $R_{\m\n}$ \cite{pirani,BS}, see \eq{Riem-delta} and \eq{Ricci-delta}. Such
singularities in $R_{\m\n}$ is not acceptable and should be
killed. As we elucidate in the appendix \ref{osj}, the condition for the
absence of the Dirac delta function is precisely the O'Brien-Synge (OS) junction
conditions. Let $\cS$ be a null surface and let it be defined by
$x^0={\rm const.}$,
the OS junction conditions require that
\be \label{OS}
g_{\m\n}\;,\quad \sum_{i,j} g^{ij}g_{ij,0}\;, \quad
\sum_{i}g^{i0}g_{ij,0}, \qquad (i,j \neq 0)
\ee
be continuous across ${\cal S}$.

The OS junction condition kills the $\d$-function singularity in
$R_{\m\n}$ and leaves us with a P.C. $R_{\m\n}$. This is in
perfect consistence with the original assumption that the stress
tensor is P.C..  The invertibility and continuity of the metric
and the piecewise continuity of $R_{\m\n}$ implies automatically
that $R$ and $R_2$ are P.C. also. Having the
$\d$-function killed, however, there is still the possibility that
$R, R_2$ may blow up at the junctions, i.e. of the form $ \sim
u^{-a}, v^{-a}, a>0$. This would be physically unacceptable and
more conditions may have to be imposed on the metric so that this
does not appear.

As for the  $\d$-function in $R_{\m\n\a\b}$ (also possibly in
$R_4$ or other higher curvature invariants), it has been proposed
\cite{pirani} to identify these discontinuity  with the impulsive
gravitational wavefront. We will give an explicit proof of this in
our appendix \ref{brink}. They are therefore physically acceptable.
However, it is also possible to
impose further conditions on the metric such that $R_{\m\n\a\b}$
and $R_4$  has no $\d$-function singularity. Similarly, one may
impose further conditions such that $R_{\m\n\a\b}$ and $R_4$  has
no poles.

Summarizing our discussion above, we conclude that  in order to
paste together the solutions  obtained from solving the
differential equation of motion in the different regions, one need
to impose the following junction condition on the metric:
\begin{itemize}
\item[{\it (1)}]  If the metric is $C^1$, then impose the Lichnerowicz
  condition. Otherwise, if the metric is piecewise $C^1$, then
impose the OS junction conditions.

\item[{\it
(2)}]  curvature invariants $R, R_2$ do not blow up at the
  junction.
\end{itemize}
In special circumstances, one may also require that
\be
\mbox{${\it (3^*)}$ $R_{\m\n\a\b}$ and  $R_4$  have no
$\d$-function singularity or
  blow up at the junction.}\;\;\;\;\;\; \nn
\ee

\section{Colliding Plane Wave Solutions with Flux}

We now apply the junction conditions to the flux-CPW solutions we
find in section 2 and use them to constraint the parameters
appearing in the solution. Let us write the near-junction
expansion of $f(u\geq 0)$ and $g(v\geq0)$  as follows:
\bea
&& f = f_0 (1- d_1 u^{n_1})\;, \quad u \sim 0^+ ,\\
&& g = g_0 (1- d_2 v^{n_2})\;, \quad v \sim 0^+.
\eea
In particular the boundary conditions will put restrictions on the
{\it boundary exponents} $n_i$.

\subsection{Imposing  junction conditions}

\underline{{\it (1)} Lichnerowicz/O'Brien-Synge junction
conditions}

First we require the metric to be continuous across the junctions.
Continuity of $e^A$ and $e^B$ is automatic. If one fixes the
normalization of the metric such that $A=B=M=0$ in the P-region,
then we get $f_0=g_0=1/2$. As for the continuity of $e^{-M}$, the
condition \eq{fg} requires
\be
\label{gcon}
b_1=1-{1\over n_1}\;, \qquad d_1=({2\over n_1})^{n_1}.
\ee
As for the condition from \eq{fga}, we just have to replace the
subscript $1$ by $2$ in \eq{gcon}.

For the solution II, there are additional constraints to continue
the metric in the L/R-region to the flat metric in the P-region,
say $e^{-M}$, $e^A$ and $e^B$ normalized to $1$, this requires
\be
a_0\cosh^{a_4}(c_1\log c_2)=1\;, \quad {\a \c^2 \over 8\d c_1^2}
\cosh^2(c_1\log c_2)=1.
\ee
These can be simplified to
\be
c_2^{1+4c_1^2 \over 2}=\cosh^2(c_1\log c_2)={8\d c_1^2 \over \a
\c^2}\;.
\ee
One can solve these  constraints for $c_1, c_2>0$ in terms of
$\gamma$
\be
c_1 = \sqrt{\frac{\a \c^2}{8 \d}}, \quad c_2=1.
\ee
Therefore we get a three-parameters family of solutions depending
on $\gamma$, $\k_1$ and $\k_2$.

Next we ask when the metric is  piecewise $C^1$. Let us consider
the junction $u=0$ in details. The analysis for the $v=0$ junction
is exactly the same. First we claim that for our solution I and
II, we have for $u\sim 0$
 \bea && U_u = \Bigl(u^{n_1-1}
\;\frac{-d_1n_1}{1+2g} + \;\mbox{l.s.t.} \Bigr)\;\Theta(u),
\label{M1} \\
&& \a V_u =\Bigl( u^{{n_1\over2}-1} \cdot \a e_1(v)
+\;\mbox{l.s.t.} \Bigr)\;\Theta(u), \label{M2} \\
&& nA_u =\Bigl(u^{{n_1\over2}-1} e_1(v) + \mbox{l.s.t.}\Bigr)\;
\Theta(u) , \label{M3}\\
&& mB_u = \Bigl(- u^{{n_1\over2}-1} e_1(v)+ \mbox{l.s.t.}\Bigr)\;
\Theta(u), \label{M4}\\
&& M_u = \Bigl(\k_1 \k_2  u^{{n_1\over2}-1} e_0(v) - \eta
u^{{n_1\over2}-1} e_1(v) + \mbox{l.s.t.}\Bigr)\;\Theta(u),
\label{M5}
\eea
where l.s.t. in the above stands for less singular terms and
$e_0(v),e_1(v) $ are  some nonzero functions of $v$. The proof can
be found in the appendix \ref{bb}. As a result of \eq{M3}-\eq{M5}, we
find that the metric is  $C^1$ if $n_1 >2$ and is  piecewise $C^1$
if $n_1 \leq 2$.

For the case that metric is $C^1$,  it is easy to see from
\eq{M3}-\eq{M5}  that it is
also at least piecewise $C^2$. Thus the Lichnerowicz condition is
satisfied. As for the case that the metric is piecewise $C^1$, i.e.
$n_1\le 2$, we need to impose the second and third OS junction
conditions which require that $U_u$ to be continuous
(i.e. equal to zero) across the junction at $u=0$. From \eq{M1},
it is easy to see the piecewise $C^1$ metric also
satisfies  the OS condition only if
\be \label{cond1}
\mbox{$1 <n_i\le 2$.}
\ee
Alternatively, we can understand why only $U_u$ is constrained by
the OS condition from the following fact: the $(n+1)$-th order
differential equation is solved provided that the boundary
conditions from the zeroth to the $n$-th derivatives are given.
Looking into EOM, we note that the only term of the second
derivative with respect to $u$ is $U_{uu}$, the others are all
terms of the first derivative with respect to $u$. Therefore,
besides the continuity of the metric at junction we need to impose
the continuity only on $U_u$ but not on $M_u$, $V_u$, $C_u$ and
$\phi_u$. Similarly for the junction condition at $v=0$.

In summary,  from imposing the Lichnerowicz or the O'Brien-Synge
junction conditions, we have the following allowed possibilities
\be \label{cond2}
 \left\{ \begin{array}{ll}
 \mbox{(i)  $1<n_i\leq 2:\;\;$} & \mbox{metric is piecewise $C^1$ }\\
\mbox{(ii) $n_i > 2$:} & \mbox{metric is at least piecewise $C^2$
},
\end{array}
\right.
\ee
in our solution I  \eq{solnX}, \eq{solnE}-\eq{solnd} and our
solution II  \eq{solnX}, \eq{solnC2}-\eq{solnd2}.

\underline{{\it (2)} on $R, R_2$}

Having imposed the above junction conditions, the Ricci tensor
$R_{\m\n}$ is at least P.C.. However it may still blow up at the
junction. Now we claim that in order for the curvature invariants
$R$ {\it and} $R_2$ not to blow up at the junction, we need the
condition $n_i \geq2$ in addition to the junction condition
imposed above. To see this, we start with
\be\label{RMAB}
R= 2 e^M R_{uv} + n e^{-A} R_{xx} + me^{-B} R_{yy},
\ee
where $R_{uv}$ etc. are given in appendix \ref{rr}. We note that only
first order derivatives of $M,A,B$ with respect to $u$ or $v$
appear in $R_{uv}, R_{xx}$ and $R_{yy}$ and there is no second
order derivatives like $A_{uu}$, therefore the singularity
behavior of $R$ is controlled by $U_u, V_u, M_u$. Now $V_u$ is the
most singular object and from  \eq{M2}, it is
\be
V_u \sim u^{{n_1\over 2}-1},
\ee
therefore $R$ is non-singular at the boundary if $n_i \geq2$.
However this condition may be too strong since there could be
cancellation among different pieces in $R$. To check whether this
is the case, one can either examine \eq{RMAB} explicitly.
Alternatively we can resort to the Einstein equations and obtains
the following simple form for $R$
\be \label{RR}
R = 2 e^M \phi_u \phi_v + \frac{m-n}{m+n} \frac{e^{M -E}}{f+g} C_u
C_v.
\ee
Using the boundary behaviour \eq{phiu} and \eq{Cu} for the flux
solution, we see that for $R$ not to blow up, one requires that $n_1 \geq 2$.

Next we consider $R_2$. It is
\be
R_2 = 2 e^{2M} R_{uv}^2 + 2  e^{2M} R_{uu} R_{vv} + n e^{-2A}
R_{xx}^2 + m e^{-2B} R_{yy}^2.
\ee
Resorting to the Einstein equations as we did before for $R$, we
find that $R_2$ involves only $\phi_u$ and $C_u$ as $R$ does. One
can easily see that $R_2$ does not blow up at the junction if the
condition $n_i \geq 2$ is satisfied.

Taking into account together with the junction condition
\eq{cond2}, we find the following physical possibilities:
\be \label{pcond1}
b_i=1-1/n_i
\ee
\be \label{pcond2}
 \left\{ \begin{array}{ll}
 \mbox{(i)  $n_i=2:\;\;$} & \mbox{metric is piecewise $C^1$ }\\
\mbox{(ii) $n_i > 2$:} & \mbox{metric is at least piecewise $C^2$
},
\end{array}
\right.
\ee
on our solution I \eq{solnX}, \eq{solnE}-\eq{solnd}, and on our
solution II \eq{solnX}, \eq{solnC2}-\eq{solnd2}.

\underline{${\it (3^*)}\; $ on $R_{\m\n\a\b}$ and  $R_4$ }

If one want, one may further restrict the solution so that
$R_{\m\n\a\b}$ and  $R_4$ have no $\d$-function singularity and do
not  blow up at the junction. Let us examine first the
$\d$-function singularity, which is possible when  the metric is
piecewise  $C^1$. We claim that for the
$\d$-function singularity in $R_{\m\n\a\b}$ or $R_4$ to disappear,
it is necessary that $n_i >2$. Thus in view of \eq{pcond2}, all
the
 piecewise $C^1$ metric of (i) results in a $\d$-function singularity in
$R_{\m\n\a\b}$ and  $R_4$.

  We have two ways to find out the condition for the absence of
the $\d$-function singularity at junction. For the general cases
one can use the result in Appendix \ref{osj} by requiring
\be
\label{Duaub}
D_{u\a u\b} \equiv - h_{\a\b} = \D(g_{\a\b,u})
\ee
and
\be
I = -4 e^{2M} [n \D(A_u) e^{-A}R_{vxvx} + m \D(B_u)
e^{-B}R_{vyvy}].
\ee
to be vanishing at $u=0$.

  For our particular flux-CPW solution I and II, it is more
straightforward to use \eq{M3}-\eq{M4} and \eq{R4c} to write down
the most singular term in $R_4$, and the result is
\be
\label{R4cc}
R_4 \sim\Bigl(u^{{n_1\over2}-1}\delta(u)+
({n_1\over2}-1)u^{{n_1\over2}-2}\Theta(u)\Bigr)\;
e_1(v)(A_{vv}-B_{vv})+{\cal O}(u^{n_1-2}\Theta(u))
\ee
where the sub-leading term $u^{n_1-2}\Theta(u)$ comes from the
terms such as $A^2_u$ and $B^2_u$ omitted in \eq{R4c}. From
\eq{R4cc} we can summarize the singularity structure in the
following table:

\begin{center}{\underline{Table I. 
Summary of the singularity structure of $R_4$ }}
\begin{tabular}{||c|c|c|c|c||}
\hline
              & $n_i=2$ & $2<n_i<4$ & $n_i=4$ & $n_i>4$
             \\\hline
Dirac-$\delta$   & yes& no  & no& no\\\hline
step       & yes & yes &  yes& no\\\hline
pole       & no & yes & no & no \\\hline
\end{tabular}
\end{center}
\bigskip
Note that the $n_1=2$ case has the sharp $\delta$-function
plus $\Theta$-function profile which corresponds to an impulsive
wavefront with a tail,
and the $n_1=4$ case has the $\Theta$-function profile, thus a
shock wavefront. For $n_1>4$ one has smooth wavefront.

Finally, we would like to comment on the necessity of imposing
pole-free condition on the curvature invariants.   The pole of the
curvature invariants is considered to be problematic because the
general relativity break downs there, for example, the black hole
singularity or the cosmological singularity at the big bang. On
the other hand, the Dirac delta function or Theta function
singularities are generally accepted and regarded as an idealized
limiting case of localized matter source, for example, in the
Khan-Penrose CPW and the shock wave considered in \cite{hooft}.
Based on these, the case $2<n_i<4$ in the Table I. cannot be
accepted. However, to allow for broader classes of CPW solutions, we
will not be so restrictive in our discussions. Instead
only the condition \eq{cond2} is imposed to ensure that $R$ and $R_2$
is not blowing up at the junction. 
However it is straightforward to also require  that 
$R_4$ not to blow up by rejecting the solutions with $2<n_i<4$.

\subsection{Physical flux-CPW solution}

Now we apply the  physical conditions \eq{pcond1}, \eq{pcond2} to
the two solutions we obtained in section 2.  Recall that
\be
b_i={\k_i^2+\e \d \over 1+a^2 \d}\;,\quad \d={m+n\over 4mn},
\ee
where $\e=1$ for the first type of solution and $\e=0$ for the
second type of solution. We  have the following
window for $\k_1^2$ and $\k_2^2$,
\be
\label{physicalk}
 {1\over2}+\d({a^2\over 2}-\e)\leq \k_1^2,\; \k_2^2
< 1+\d(a^2-\e)\;.
\ee
Here, the metric is piecewise $C^1$ when the equality sign holds, and in this
case the metric satisfies the OS junction condition. Otherwise the metric
is piecewise $C^1$ and satisfies the Lichnerowicz junction condition.

To discuss further,
let us divide the allowed solutions into non-dilatonic and dilatonic cases.

\begin{itemize}
\item Case 1. If there is no dilaton so that $a=0$, $\k_1=\k_2=0$, then
$b_i= \e\d$.

In this case, using $\d \leq 1/2$, it is easy to see that only
$m=n=1$ and $a=0$ (i.e. no dilaton) of the type I solution  is
allowed. And the metric is piecewise $C^1$ with $n_i =2$. This is
precisely the  original case of
Bell-Szekeres \cite{BS}.  Note that from table I, this
solution is impulsive, i.e. has a $\d$-function profile in
$R_{\m\n\a\b}$ and  $R_4$.

The higher dimensional generalization  of the 4-dimensional
Bell-Szekeres solution with a $n$-form potential (by which we mean
$m=n$ and non-dilatonic) has been considered recently  by Gutperle
and Pioline in \cite{GP}. They found for their solution
$n_1=2n/(2n-1)<2$ for $n>1$ and that $R_2$ blows up at the
junction.  This is consistent with what our analysis.

The higher dimensional Einstein-Maxwell CPW
solutions (i.e. $a=0$ and $m> n=1$) have been considered in
\cite{higherBS}. In terms of our notation,
their solution are our type I solution with
$4/3 <n_i = 4m/(3m -1) \leq 8/5$ since  $m \geq 2$.
Therefore their metric is piecewise $C^1$ and satisfies the OS
condition. However $R$ and $R_2$ blows up.

\item Case 2. There is a dilaton profile.

Note that the bound at the LHS of \eq{physicalk} is nonnegative and that
the size of the $(\k_1, \k_2)$ window does not depend on the flux
amplitude $\c$ in the solution II.  We thus have a 2 (or 3)
parameters family of solutions labeled by $\k_1$, $\k_2$ (and
$\c$). A 4-dimensional solution has been considered  by Gurses and
Sermutlu in \cite{Gurses1} where $m=n=1$, $\e=1$, and $|\k_1|=
|\k_2| = a/ 2$, so that $n_1=n_2=2$.  The metric of this solution
is piecewise $C^1$. Our type I solutions give generalization of this
solution.  Moreover our type II  solution  is
completely new and has never appeared in the literature.

\end{itemize}

\subsection{Future singularity  of the solution}

As known that in 4-dimensional spacetime the future curvature
singularity is a general outcome of the collisions of the
gravitational plane waves even with arbitrarily small density
\cite{Tipler}. It is then curious to see if the future curvature
singularity will also  generically appear in our new higher
dimensional flux-CPW solutions.
We will investigate this issue in this subsection.

We first note that the metric may blow up or vanish at
\be
\label{f+g}
f(u)+g(v)=0,
\ee
which define a hypersurface ${\cal S}_0$. Near ${\cal S}_0$ we
then have
\be
e^E \sim (f+g)^{-1}\;, \quad e^X \sim (f+g)^{\k_1+\k_2}\;,\quad
\cosh(c_1\log {c_2\over f+g})\sim (f+g)^{-c_1}.
\ee
This results in the following singular behavior near $f+g=0$
\bea
\label{Em}
e^{-M}&\sim & (f+g)^{-b_3-{\eta\over
\a}(1-a(\k_1+\k_2))+(1-\e)({\eta\over \a}-a_4c_1)}\;,
\\
\label{Ea}
e^{nA}&\sim& (f+g)^{1-{1\over \a}(1-a(\k_1+\k_2))+(1-\e)({2\over
\a})(1-c_1)}\;,
\\
\label{Eb}
e^{mB}&\sim&(f+g)^{1+{1\over \a}(1-a(\k_1+\k_2))-(1-\e)({2\over
\a})(1-c_1)}\;,
\\
\label{Ephi}
e^{\phi}&\sim& (f+g)^{{1\over \a}(\d a+\k_1+\k_2)-(1-\e)({2\d
a\over \a})(1-c_1)}\;,
\\
C_u, C_v &\sim& 1\;,
\eea
where $\e= 1$ (resp. 0) for the type I (resp. II) solution as
before. The regularity and the invertibility of the metric require
that the exponents in \eq{Em}, \eq{Ea}, \eq{Eb} vanish. However,
it is easy to see that it is impossible for the exponents of
\eq{Ea} and \eq{Eb} to vanish at the same time. Therefore we
conclude that at $f+g=0$ the metric is singular, i.e. either blows
up or vanishes.
In particular we have a Killing horizon when $e^A$ or $e^B$
vanishes. For example, the Bell-Szekeres solution has a Killing
horizon at $f+g=0$.

 The above metric singularity could be just a coordinate singularity
if the curvature invariants do not blow up on ${\cal S}_0$. Next
we check the curvature singularity which may appear in $R$ , $R_2$
and $R_4$.  First we note that, from \eq{Em}-\eq{Eb} we have
\be
{\partial^{\ell_1+\ell_2} M \over \partial u^{\ell_1}
\partial v^{\ell_2}}\sim {\partial^{\ell_1+\ell_2} A \over \partial u^{\ell_1}
\partial v^{\ell_2}}\sim {\partial^{\ell_1+\ell_2} B \over \partial u^{\ell_1}
\partial v^{\ell_2}} \sim (f+g)^{-(\ell_1+\ell_2)}
\ee
near ${\cal S}_0$.  Then from the expressions of the Ricci tensor
and Riemann tensor listed in the Appendix \ref{rr}, it is easy to check
that the most singular terms near ${\cal S}_0$ in $R^2$, $R_2$ and
$R_4$ are all taking the following generic form \footnote{ This is
not the case when there is no dilaton. In this case the only
solution is the Bell-Szekeres solution and there $R=R_2=0$ and
$R_4= const$.
As mentioned above, the singularity at $\cS_0$ is not a curvature
singularity, but that of a Killing horizon. However when the
dilaton is tuned on, one can easily check that the coefficient of
\eq{fs} in  $R^2$, $R_2$ and $R_4$ are nontrivial function of, say
$u$, on the hypersurface ${\cal S}_0$. }
\be
\label{fs}
e^{2M}(f+g)^{-4} \sim [(f+g)^{b_3+{\eta\over
\a}(1-a(\k_1+\k_2))-(1-\e)({\eta\over \a}-a_4c_1)-2}]^2 \;.
\ee
Therefore, to avoid the future curvature singularity on ${\cal
S}_0$ it is required that the exponent in \eq{fs} to be
non-negative, or equivalently
\be \label{Rcc1}
(\k_1+\k_2)^2+\eta a (\k_1+\k_2) +\d - \eta +\a(1+\d)\le
-2(1-\e)(1-c_1)(\eta-2\d c_1),
\ee
where we recall that
\be
b_3=1-\d-{\d+(\k_1+\k_2)^2 \over \a}-(1-\e)(\frac{\eta +4 \d
c_1^2}{\a})\;, \quad a_4={4\d \over \a}+{2\eta \over \a}
\ee
and
\be
\d={n+m \over 4mn}\;, \quad \a=1+a^2\d\;,\quad \eta={m-n\over
2mn}\;.
\ee
Note that the condition \eq{Rcc1} makes the metric component
$e^{-M}$ blows up on ${\cal S}_0$.

  The condition \eq{Rcc1} is a complicated one, and in general it
can be violated so that a curvature singularity will develop at
the late time. To see this, we first note that the LHS of
\eq{Rcc1} is always greater than 1,
\be
{\rm LHS} = (\k_1+\k_2+ \frac{\eta a}{2})^2 + 1+ a^2 \d +
(2\d-\eta) + a^2(\d^2 - \frac{\eta^2}{4} ) >1
\ee
for any $\k_1$ and $\k_2$. Therefore type I solution will always
develop late time curvature singularity. As for type II solution,
we note that the RHS of \eq{Rcc1} is
\be
{\rm RHS} = -4\d (c_1 - \frac{1}{4n\d})^2 + \frac{n}{m(n+m)} <1 .
\ee
Thus \eq{Rcc1} can never be satisfied and curvature singularity
will always develop. Therefore, in conclusion, we find that both
the type I and type II solutions will develop curvature
singularity in the future hypersurface $f+g=0$.

\section{Conclusions and Discussions}

   In this paper we have constructed  physical solutions of CPW
in string theory with non-zero form flux. These solutions solve
the EOM in the interior of each region. Moreover: {\it (1)} they
satisfy the  Lichnerowicz/O'Brien-Synge junction conditions, and
{\it (2)} the curvature invariants $R$, $R_2$ do not blow up at
the junctions. The results of this paper can be summarized in the
Table.

\begin{center}{\underline{Table II.  Physical flux-CPW  solutions }}
\begin{tabular}{||c|p{2.95cm}|p{2.55cm}||p{2.95cm}|p{2.95cm}||}
\hline \multicolumn{5}{|c|}{Summary of the CPW solutions with
$(n+1)$-form flux in $m+n+2$ dimensions }
 \\ \hline\hline
   &  \multicolumn{2}{c||}{without dilaton} & \multicolumn{2}{c||}{with nonzero
dilaton} \\
   \hline
   &  \quad Soln. I ($\e=1$)  & Soln. II ($\e=0$) & \quad Soln. I ($\e=1$)
&\quad  Soln. II ($\e=0$) \\ \hline
  $m=n$ & only $m=n=1$, i.e., Bell-Szekeres & no physical solution
&\eq{solnX}, \qquad \eq{solnE}-\eq{solnd} &
 \eq{solnX}, \qquad \eq{solnC2}-\eq{solnd2}\\
  \cline{1-3}
  $m\ne n$ &  no physical solution & no physical solution &
parameterized by $\k_1$, $\k_2$ subjected to \eq{physicalk}  &
  parameterized by $\k_1$, $\k_2$, $\gamma$ subjected to \eq{physicalk} \\
  \hline
\end{tabular}
\end{center}

\bigskip

In this paper, we have obtained two types of solution using two
 different ansatz.
 The solution I follows  the ansatz of the
original Bell-Szekeres solution. The solution II is new and has
never appeared in the literature before. We see that the original
Bell-Szekeres solution cannot be generalized to higher dimensions
unless the dilaton field is turned on. Roughly speaking, a
particular combination \eq{EX} of the dilaton field and the metric
component $V$ is given by \eq{solnX}  and behaves like the $V$ in
the pure CPW. As we explained before, this field $X$ makes an
important contribution to the metric, and when suitably restricted
(see \eq{physicalk}) could smoothen out the singularities in the
solution  and lead to physically acceptable CPW solution.

We have also shown that all the solutions (except for the
Bell-Szekeres solution as explained before) in the
Table II.
will result in a late time curvature singularity on a hypersurface
${\cal S}_0$ defined by \eq{f+g}. In \cite{Tipler} Tipler showed
that the collisions of 4-dimensional gravitational plane waves
will develop a late time curvature singularity due to the
artificial plane symmetry.  Our results support the generalized
version of Tipler's theorem in higher dimensional supergravity.

One of our original motivation in this project was to construct
flux-CPW in M-theory since the spectrum of fields are very simple.
However we failed to obtain such solution, both for the 11
dimensional and for the 27 dimensional M-theory \cite{hs}. In
retrospect, this is understandable since there is no dilaton in
these theories. A more general ansatz than those we considered may
be needed. So far only purely gravitational CPW solution has been
constructed in these theories. One may wonder whether the absence
of flux-CPW solution in these theories has any fundamental
meaning? It is important to try to construct flux-CPW solutions
for these non-dilatonic theories.

It may be interesting to consider CPW with more than one higher
form field turning on. For example, one may consider having a
$F_{[n+1]}$ field strength together with a  a $F_{[m+1]}$ field
strength simultaneously. This will be very easy to deal with using
our ansatz \eq{metric-ansatz} for the metric. A more general
metric ansatz will be needed when the potential do not have
complementary dimensions. Similarly, it may be  interesting to
study the scattering of waves with both the dilaton and axion
turned on in IIB string theory.

Finally, we hope that the physical flux-CPW solutions 
constructed in this paper
will shed new lights or find applications on the issues of
cosmological singularity or its resolution in the context of
string theory.

\section*{Acknowledgements}
We would like to thank Takeo Inami for interesting discussions and
useful comments.  BC would like to thank the hospitality of NCTS
during his visit. CSC would like to thank Michael Gutperle for
interesting conversations. FLL would like to thank the warm
hospitality of ICTS and ITP at Beijing while this work was in
progress and also to thank Miao Li for discussions. We acknowledge
grants from Nuffield foundation of UK, NCTS and NSC of Taiwan.

\appendix

\section{Ricci and Riemann tensors} \label{rr}
For the metric
\be
\label{appen1}
ds^2 = 2 e^{-M} dudv + e^A \sum_{i=1}^n dx_i^2 +  e^B \sum_{j=1}^m
dy_j^2
\ee
with  the functions $M, A, B$ being functions of $u,v$. We have
\bea
&& R_{uu} = -\frac{1}{2} [nA_{uu} +m B_{uu} + nM_u A_u + mM_u B_u
+ \frac{1}{2}(n A_u^2 + m
B_u^2)], \\
&& R_{vv} = -\frac{1}{2} [ nA_{vv} +m B_{vv} + nM_v A_v + mM_v B_v
+ \frac{1}{2}(n A_v^2 + m
B_v^2) ], \\
&&R_{uv}= M_{uv} - \frac{n}{2}A_{uv} - \frac{m}{2}B_{uv} -
\frac{1}{4}(nA_uA_v+mB_uB_v) , \\
&& R_{x x} = -\frac{1}{2}e^{M+A}
[2A_{uv} +n A_u A_v +\frac{m}{2}(A_u B_v +A_v B_u) ], \\
&& R_{yy} = -\frac{1}{2}e^{M+B} [ 2B_{uv} +m B_u B_v +
\frac{n}{2}(A_u B_v +A_v B_u)].
\eea
Here $x= x_i$ for  $i=1, \cdots, n$ and  $y= y_j$ for  $ j=1,
\cdots, m$.

In the following we list the independent nonvanishing components
of the Riemann tensor for the metric \eq{appen1}:
\bea
R_{uvuv}&=&-e^{-M}M_{uv}\;,
\\
R_{xyxy}&=&{-1\over 4} e^{M+A+B}(A_uB_v+A_vB_u)\;,
\\
R_{uxvx}&=&-e^A({1\over2} A_{uv}+{1\over4}A_u A_v)\;,
\\
R_{vxvx}&=&-e^A({1\over2} A_{vv}+{1\over2} M_v A_v +{1\over 4}
A^2_v)\;,
\\
R_{uxux}&=&-e^A({1\over2} A_{uu}+{1\over2} M_u A_u +{1\over4}
A^2_u)\;,
\\
R_{uyvy}&=&-e^B({1\over2} B_{uv}+{1\over4} B_u B_v)\;,
\\
R_{vyvy}&=&-e^B({1\over2}B_{vv}+{1\over2} M_vB_v +{1\over4}
B_v^2)\;,
\\
R_{uyuy}&=&-e^B({1\over2} B_{uu}+{1\over2} M_uB_u +{1\over4}
B^2_u)\;.
\eea
As usual
\be
R_{\m\n\r\s}=R_{[\m\n][\r\s]}=R_{[\r\s][\m\n]}\;.
\ee
Moreover, we also have that near $u=0$ or $v=0$,
\be
\label{R4c}
R_4 = {e^{2M}\over 4}(n A_{uu}A_{vv}+m B_{uu}B_{vv})+\cdots\; ,
\ee
where $\cdots$ denotes terms that contains  first order
derivatives with respect to $u$ or $v$ and are less singular at
the junctions than the second derivative terms $A_{uu}$ etc.

\section{On the O'Brien-Synge Junction Conditions and Beyond} \label{osj}

As noted by  Bell-Szekeres \cite{BS},
the original Khan-Penrose pure CPW metric is piecewise $C^1$ and
does not satisfy the Lichnerowicz condition. However the  Khan-Penrose
solution is physical and perfectly acceptable. Moreover, in the case of
electromagnetism  where there could be different electromagnetic
field configurations in different regions of spacetime, it is
necessary to allow for metrics that are piecewise $C^1$
such that the Ricci tensor is P.C..
However in general, the Ricci tensor of a piecewise $C^1$ metric has
$\d$-function  singularity. In this appendix,
we  study the conditions for the absence of such
singularities. We will show explicitly that the requirement that {\it the
Ricci tensor to be P.C. is equivalent to the  O'Brien-Synge
junction conditions on the metric}. Given a null \footnote{ We
recall here the reason  why  null surface discontinuity is
relevant for the problem of CPW with P.C. higher form field. In
general, consider a $n$-form field strength $F_{[n]}$, and assume
that  $F$ be P.C. with  a discontinuity across a certain
hypersurface $\cS$ described by the equation $u(x^\m)=0$. Then
$F_{[n]}$ takes the form
\be
F_{\m_1 \m_2 \cdots \m_n}  = f_{\m_1 \m_2 \cdots \m_n} +
\psi_{\m_1 \m_2 \cdots \m_n} \Th(u),
\ee
where $f_{\m_1 \m_2 \cdots \m_n}$ is continuous. Then
\be
F_{\m_1 \m_2 \cdots \m_n, \r} =  f_{\m_1 \m_2 \cdots \m_n, \r} +
\psi_{\m_1 \m_2 \cdots \m_n, \r} \Th(u) + \psi_{\m_1 \m_2 \cdots
\m_n} u_\r \d(u),
\ee
where $u_{\rho} := \del_\r u$ is the normal derivative to $\cS$.
It follows from the Bianichi identity and the field equation that
\be
\psi_{[\m_1 \m_2 \cdots \m_n} u_{\r]} =0, \quad \psi_{\m_1 \m_2
\cdots
  \m_{n-1} \r } u_\r =0.
\ee
Contracting the first equation with $u^\r$ and we get $u_\r u^\r
=0$. Hence the surface of discontinuity must be null. } surface
${\cal S}$  defined by $x^0=const.$,
the OS junction conditions require that
\be
g_{\m\n}\;,\quad \sum_{i,j} g^{ij}g_{ij,0}\;, \quad
\sum_{i}g^{i0}g_{ij,0}, \quad i,j \neq 0
\ee
be continuous across ${\cal S}$.

To prove this, we first recall that that if the metric $g_{\m\n}$
is continuous and $g_{\m\n,\rho}$ is P.C. across ${\cal S}$
defined by $u(x^\m)=0$, then $g_{\m\n,\rho}$ must take the form
\cite{pirani}
\be
\label{g1}
g_{\m\n,\rho}={g}^{(0)}_{\m\n,\rho}+h_{\m\n}u_{\rho}\Theta(u)
\ee
for some $h_{\m\n}$ and $u_{\rho}$ is the normal to ${\cal S}$.
The piece ${g}^{(0)}_{\m\n,\rho}$ is continuous across ${\cal S}$,
and $\Theta(u)$ is the Heaviside step function. Furthermore,
Bell-Szekeres \cite{BS} showed that \eq{g1} yields
\bea \label{Riem-delta}
R_{\m\n\r\s}&=&R^{(1)}_{\m\n\r\s}+2u_{[\m}h_{\n][\r}u_{\s]}\delta(u)\;,
\\
g_{\m\n,\r\s}&=&g^{(1)}_{\m\n,\r\s}+h_{\m\n}u_{\r}u_{\s}
\delta(u)\;, \nn
\eea
where the superscript $(1)$ refers to the piecewise continuity of
the quantity across ${\cal S}$. It follows immediately that
\be \label{Ricci-delta}
R_{\n\s} = R_{\n\s}^{(1)}+ 2 N_{\n\s}\d(u), \quad
\mbox{where}\quad N_{\n\s}:= h_{\m\n}u^{\m} u_\s + h_{\m\s}u^{\m}
u_\n- h^{\l}_{\l} u_\n u_\s.
\ee
Therefore for the Ricci tensor to be P.C., i.e. the $\delta(u)$
piece vanishes, it is necessary that
\be\label{BS}
N_{\n\s}= h_{\m\n}u^{\m} u_\s + h_{\m\s}u^{\m} u_\n- h^{\l}_{\l}
u_\n u_\s =0 .
\ee

 Now we show that the condition \eq{BS} for $R_{\m\n}$ to be P.C. is
equivalent to OS conditions mentioned above. Take $u=x^0 =const.$
to be the null surface $\cS$. And choose the normal to $\cS$ to be
$u_\m = \d^\m_0$. That the surface is null implies that  $g^{00}
=0$.  The condition \eq{BS} gives the following nontrivial
conditions
\be
N_{00} = 2 h_{\m 0} g^{\m 0} - h^\l_\l =0,\quad \mbox{and}\quad
N_{0i}=  h_{i \m} g^{\m 0} =0. \label{n}
\ee
Denotes  by $\D(H)$ the discontinuity of a function $H$ across
${\cal S}$, it is $h_{\m\n} = \D(g_{\m\n,0})$. We have
\be
N_{00} = 2 \D(g_{\m 0,0})g^{\m 0} -  \D(g_{\m\l,0})g^{\m\l} = -
\D(g_{i j,0})g^{i j}
\ee
and
\be
N_{0i} = \D(g_{i j,0})  g^{j 0}.
\ee
Therefore the condition \eq{n} is precisely the same as the OS
junction condition. This conclude our proof.

We remark that for our metric ansatz \eq{metric-ansatz}, the OS
junction condition requires that, say, across the junction $u=0$:
\be
\mbox{$U,V, M$ continuous}, \qquad \mbox{and} \qquad U_u \to 0, \quad u \to 0^+.
\ee
Note that there is no requirement on the other normal derivatives
$V_u$ or $M_u$.

Next we examine the singularity in $R_{\m\n\r\s}$ and $R_4$. Let
us denote the coefficient of the $\d$-function singularity in
\eq{Riem-delta} by
\be
D_{\m\n\r\s}:= 2 u_{[\m}h_{\n][\r}u_{\s]}\;.
\ee
Consider the junction $u=x^0= 0$, it is easy to see that the only
independent nonvanishing  component of $D_{\m\n\r\s}$ is
\be \label{van}
D_{0\a0\b} = - 2 h_{\a\b} = -2 \D(g_{\a\b,0}).
\ee
Thus the $\d$-function singularity in $R_{\m\n\r\s}$ is absent if
the normal derivatives of the metric is continuous across the
junction.

As for $R_4$, we note that for the form of our metric ansatz
\eq{metric-ansatz}, it is easy to show that
\be \label{I}
R_4 =R_4^{(1)} + I \d(u), \quad I := 4 D_{0\a0\b} R^{0\a0\b}
\ee
across the $u=0$ boundary.  Here $R_4^{(1)}$ is P.C.. Note that
there is no $(\d(u))^2 $ term.

\section{CPW solution without flux} \label{cpw-fluxless}

In this appendix, we give  the pure colliding gravitational and
dilatonic plane waves in the higher dimensional dilaton gravity
{\it without} turning on the form flux. Note that both the dilaton
$\phi$ and the metric component $V$ obey the same equation as in the standard
purely gravitational case, see \eq{e2} and \eq{V} by setting $C$
and $a$ to zero. Explicitly, the pure dilatonic and  gravitational
CPW solution is given by
\bea
\phi&=&\k_1 \log{w-p \over w+p} +\k_2 \log{r-q\over r+q}\;,\\
V&=&k_1 \log{w-p \over w+p} +k_2 \log{r-q\over r+q}\;,\\
e^{-M}&=&f_ug_v [(1-2f)(1+2g)]^{-b_1}[(1+2f)(1-2g)]^{-b_2}
(f+g)^{-b_3}
\nn\\
&\cdot & [{1\over2}+2 f g + 2 pqrw]^{-2\k_1 \k_2-2\d k_1 k_2} \;
\Bigl[({w-p\over w+p})^{k_1}({r-q\over r+q})^{k_2}\Bigr]^{\eta}\;,
\\
e^{nA}&=& (f+g)\; ({w-p\over w+p})^{k_1}({r-q\over
r+q})^{k_2}\;, \label{solnA3}\\
e^{mB}&=& (f+g)\; \Bigl[({w-p\over w+p})^{k_1}({r-q\over
r+q})^{k_2}\Bigr]^{-1}\;,
\eea
where
\be
b_1:=\k_1^2+\d k_1^2\;,\quad b_2:=\k_2^2+\d k_2^2\;,\quad
b_3:=1-\d-(\k_1+\k_2)^2-\d(k_1+k_2)^2\;,
\ee
and $\d$ and $\eta$ are defined as before.

  As for the junction condition to be imposed for these solutions, we
have already analyzed that one should impose \eq{pcond2}. This gives
\be {1\over 2}\leq \k_1^2+\d k_1^2,\; \k_2^2+\d k_2^2<1\;,
\ee
where equality holds when the metric is piecewise $C^1$.
The parameters window is quite different from \eq{physicalk} for
the flux-CPW.

The 4-dimensional Khan-Penrose solution has $n_1=2$, the metric is
piecewise $C^1$ and satisfies the OS condition. The
higher dimensional generalization of the 4-dimensional
Khan-Penrose solution (which we mean a scattering of purely
gravitational wave with $m=n$, and where the metric is piecewise $C^1$)
can be obtained by setting $\k_1=\k_2=0$ in the above, and the
$b_i=1/2$ condition gives
\be
k_i^2=m=n.
\ee
This is in contrast to the 4-dimensional Bell-Szekeres solution
which has no higher dimensional generalization satisfying the OS
junction condition.

Finally we show that singularity always develop in the F-region at
the hypersurface $\cS_0: f+g=0$.  As in section 4.3, it is easy to
see that the most singular terms near  $f+g=0$ in $R^2$, $R_2$ and
$R_4$ are all taking the same generic form
\be
\label{fs1}
e^{2M}(f+g)^{-4} \sim [(f+g)^{b_3 -\eta (k_1+k_2)-2}]^2.
\ee
Future curvature singularity can be avoid if the exponent is
non-negative, i.e. if
\be
0\geq 1+ \d + (\k_1+\k_2)^2 +\d(k_1+k_2)^2 + \eta (k_1+k_2).
\ee
However RHS is equal to
\be
{\rm RHS} = (\k_1+\k_2)^2 +\d(k_1+k_2 + \frac{\eta}{2\d})^2 + 1+
\frac{1}{\d}(\d^2-\frac{\eta^2}{4}) \geq 1.
\ee
Hence future curvature singularity will always develop.

\section{Boundary behaviour of the CPW solution} \label{bb}

In this appendix, we will analysis the boundary behaviour of the
CPW solution.

\underline{Flux-CPW}

We claim that for our flux-CPW solution of type I and type II, we
have as $u \sim 0^+$,
 \bea && U_u = \Bigl(u^{n_1-1}
\;\frac{-d_1n_1}{1+2g} + \;\mbox{l.s.t.} \Bigr)\;\Theta(u),
\label{Ma1} \\
&& \a V_u =\Bigl( u^{{n_1\over2}-1} \cdot \a e_1(v)
+\;\mbox{l.s.t.} \Bigr)\;\Theta(u), \label{Ma2} \\
&& nA_u =\Bigl(u^{{n_1\over2}-1} e_1(v) + \mbox{l.s.t.}\Bigr)\;
\Theta(u) , \label{Ma3}\\
&& mB_u = \Bigl(- u^{{n_1\over2}-1} e_1(v)+ \mbox{l.s.t.}\Bigr)\;
\Theta(u), \label{Ma4}\\
&& M_u = \Bigl(\k_1 \k_2  u^{{n_1\over2}-1} e_0(v) - \eta
u^{{n_1\over2}-1} e_1(v) + \mbox{l.s.t.}\Bigr)\;\Theta(u),
\label{Ma5}
\eea
where l.s.t. in the above stands for less singular terms and
$e_0(v),e_1(v) $ are  some nonzero functions of $v$.

To prove this, we start with
\be \label{Xu}
X_u = {f_u\over f+g}\frac{\k_1 w r + \k_2 p q}{\sqrt{{1\over 4}
-f^2}} = - u^{{n_1\over2}-1} \cdot  \k_1 n_1 \sqrt{d_1}\sqrt{1
\over 1+2g}  + \;\mbox{l.s.t.}
\ee
for $u\sim 0^+$. And for the solution I,
 \be \label{Eu}
E_u^{\rm (I)} = -{f_u\over f+g} \sqrt{1-4g^2\over 1-4f^2} =
u^{{n_1\over2}-1} \cdot n_1 \sqrt{d_1\over 2}\sqrt{1 - 2g\over
1+2g} + \;\mbox{l.s.t.},
\ee
which has the same singular behavior as $X_u$; however, for the
solution II,
\be
\nn E_u^{\rm (II)}={f_u\over f+g}(1-2c_1 \tanh(c_1\log{c_2\over
f+g})),
\ee
which is less singular than $X_u$. Combining these together, we
have
\be \label{Vu}
\a V_u = E_u +a X_u = u^{{n_1\over2}-1} \cdot \a e_1(v)
+\;\mbox{l.s.t.},
\ee
for some nonzero function $e_1(v)$. Now since
\be \label{Uu}
U_u=f_u/(f+g) \sim u^{n_1-1}/(\frac{1}{2}+g),
\ee
is less singular compared to $V_u$. Therefore from \eq{UV}, we
have $n A_u  \sim - m B_u \sim V_u $, i.e.
\eq{M3}, \eq{M4}.
As for $M_u$, we have for $u\sim 0^+$
\be
S_u = {2b_1 f_u \over 1-2f} + \frac{\k_1 \k_2 n_1\sqrt{d_1/2}}{\a}
u^{{n_1\over2}-1} \sqrt{1 - 2g\over 1+2g} +  \;\mbox{l.s.t.} =
{2b_1 f_u \over 1-2f} + \k_1 \k_2 u^{{n_1\over2}-1} e_0(v) +
\;\mbox{l.s.t.}  ,
\ee
where $e_0(v)$ is  some nonzero function. Therefore, from \eq{SM},
\eq{SolnSI}, \eq{SolnSII}, we have the following expansion of $M_u$
\bea \label{Mu0}
M_u &=&-{f_{uu}\over f_u}+ S_u - \eta V_u +
\mbox{l.s.t.}\nn \\
&=& -{n_1-1-b_1n_1 \over u}\; + \k_1 \k_2 u^{{n_1\over2}-1} e_0(v)
- \eta u^{{n_1\over2}-1}\cdot e_1(v) + \;\mbox{l.s.t.}\;.
\label{Mu} \eea
The first term in \eq{Mu} is zero because $b_1=1-1/n_1$. And we arrive at
\eq{M5}.
For convenience, we also note that the singular behavior of $\a
\phi_u=X_u-a\d E_u$ is the same as $V_u$ \be \label{phiu} \phi_u
\sim u^{{n_1\over 2}-1}, \quad u \sim 0^+ . \ee Note also the
following behaviour of the flux potential \bea \label{Cu} &&
C_u^{\rm (I)} = -\frac{\c}{2}(wr+pq) \frac{f_u}{\sqrt{{1\over
4}-f^2}} \sim u^{{n_1\over2}-1}  , \quad \mbox{for the solution
I,}
\;\;\;\;\;\;\;\;\; \\
&& C_u^{\rm (II)} = \c f_u\sim u^{n_1-1}  ,
\quad\mbox{for the solution II.}
\eea

\underline{CPW without flux}

For the CPW solution in appendix \ref{cpw-fluxless},
it is easy to see that as $u \sim 0^+$:
\bea
&& U_u, V_u, A_u, B_u \sim\Bigl( u^{n_1-1}  + \;\mbox{l.s.t.}
\Bigr)\;\Theta(u), \label{Mb1} \\
&& M_u = \Bigl( (\k_1 \k_2+ \d k_1 k_2)  u^{{n_1\over2}-1} \tilde{e}_0(v) -
\eta u^{n_1-1} \tilde{e}_1(v) + \mbox{l.s.t.}\Bigr)\;\Theta(u),
\label{Mb2}
\eea
where l.s.t. in the above stands for less singular terms and
$\tilde{e}_0(v),\tilde{e}_1(v) $ are  some nonzero functions of $v$.

\section{Incoming wave in the Brinkmann coordinates and Impulsive
  wavefront} \label{brink}

In the literature, the CPW solution is usually obtained using the
metric ansatz of the form \eq{metric-ansatz}, i.e. in the Rosen
coordinates. Sometime it is also useful to rewrite the solution in
the Brinkmann coordinates. In this appendix we will show that
$\d$-function singularity appearing in the Riemann tensor
$R_{\m\n\r\s}$ can be identified with the impulsive component in
the wavefront of the wave when written in the Brinkmann
coordinates.

In the incoming region, e.g., in the R-region ($v<0$), the metric
in  the Brinkmann coordinates takes the form
\be
\label{brinkman}
ds^2 = 2 dx^+ dx^- + \Bigl(H_x(x^+)\sum_{i=1}^n X_i^2 +
H_y(x^+)\sum_{j=1}^m Y_j^2\Bigr)(dx^+)^2 + \sum_{i=1}^n dX_i^2 +
\sum_{j=1}^m dY_j^2 ,
\ee
where $x^+$ is related to $u$ through the relation
\be
\label{EE1}
e^{-M} du = d x^+.
\ee
Note that this relation tells us that $x^+$ is monotonically increasing with
respect to $u$. Without loss of generality, we can pick $x^+ =0$ to 
correspond to $u=0$.  
Thus one can replace $\Theta(u)$ by
$\Theta(x^+)$ and $\delta(u)$ by $\delta(x^+)$.

The metric in the Brinkmann coordinate is related to that in the Rosen
coordinate  by
\bea
&& H_x = e^{-A} \frac{d^2
e^A}{dx^{+2}}=e^{2M}(A_{uu}+M_uA_u+A_u^2),
\\ && H_y = e^{-B}
\frac{d^2 e^B}{dx^{+2}}=e^{2M}(B_{uu}+M_uB_u+B_u^2).
\eea
They contains the similar second derivative terms as in $R_4$ of
\eq{R4c} so that their near junction behaviors are the same as the
one of $R_4$, which is summarized in Table I. We then conclude
that the singular structure of $R_4$ at the junction is the same
as the wavefront profile of the wave when written in the Brinkmann
coordinates.

 To be complete and explicit, we can write down the form of the
Brinkmann wavefront profile by using the near junction behavior
of $A_u$, $B_u$ and $M_u$ listed in \eq{M3}-\eq{M5}. Near $u=0$, we  get
\bea
H_x&=&{e_1\over n}
\Bigl[u^{{n_1\over2}-1}\delta(x^+)+
({n_1\over2}-1)u^{{n_1\over2}-2}\Theta(x^+)+{e_1\over n}
(1-n\eta)u^{n_1-2}\Theta(x^+)\Bigr] + \;\mbox{l.s.t.},
\\
H_y&=&{e_1\over m}
\Bigr[u^{{n_1\over2}-1}\delta(x^+)+
({n_1\over2}-1)u^{{n_1\over2}-2}\Theta(x^+)+{e_1\over m}
(1-m\eta)u^{n_1-2}\Theta(x^+)\Bigr]+ \;\mbox{l.s.t.} ,\;\;\;\;\;\;
\eea
where have set $v=0$ in \eq{M3}-\eq{M5},
used the fact $e_0(0)=0$ and abbreviated  $e_1(0):=e_1$. 
As expected, their singular structures are the same as
the one given in \eq{R4cc}.

\ed

 In general the metric in the R-region and P-region are some
functions of $f$
\be
e^A = F(f(u \Th(u) )), \quad e^B = G(f(u\Th(u))), \quad e^M =
L(f(u\Th(u))),
\ee
where the argument $u\Th(u)$ take care of the extension to cover
the P-region. Using the above, one can easily obtain
\be
H_x = \frac{L}{F}(L_u F_u + L F_{uu}), \quad H_y = \frac{L}{G}(L_u
G_u + L G_{uu}).
\ee
The double derivatives
\be \label{Fuu}
F_{uu} = F' f_{uu} + F'' f_u^2
\ee
($'$ denotes derivative with respect to $f$) contains the $f_{uu}$
term and contributes a $\d(u)$-term  to $H_x$. $F_u=F' f_u$, $L_u
=L' f_u$, as well as  the second term in \eq{Fuu} depends on $f_u$
and contribute a $\Th(u)$ term to $H_x$. Collecting all these
terms,  we see that near the junction $x^+=0$, $H_x$ is a sum of a
$\d$-function and a $\Th$-function
\be
H_x =  D_1 \d(x^+) + D_2(x^+) \Th(u(x^+)),
\ee
where
\be
D_1:= \frac{L^2 F'}{F}\Bigl|_{f=f(0)=1/2} \cdot f_{uu}(0)
\qquad D_2(u):= \frac{L(L F')'}{F} f_u^2 \Bigl|_{f(u)} \;.
\ee
\begin{thebibliography}{99}


\bibitem{sz}
P.~Szekeres, ``Colliding Gravitational Waves,'' Nature  {\bf 228}
(1970) 1183.

\bibitem{KP} K.~A.~Khan and R.~Penrose, ``Scattering of two
impulsive gravitational plane waves,'', Nature {\bf 229} (1971)
185.

\bibitem{griffiths}
J.~B.~Griffiths, ``Colliding Plane Waves In General Relativity,''
Clarendon Press 1991, Oxford.

\bibitem{belinski}
V. Belinski, E. Verdaguer, ``Gravitational Solitons'', Cambridge
University Press, 2001.

\bibitem{Tipler}
F.~J.~Tipler, ``Singularities From Colliding Plane Gravitational
Waves,'' Phys.\ Rev.\ D {\bf 22} (1980) 2929.

\bibitem{yu88}
U.~Yurtsever, ``Colliding Almost Plane Gravitational Waves:
Colliding Plane Waves And General Properties Of Almost Plane Wave
Space-Times,'' Phys.\ Rev.\ D {\bf 37} (1988) 2803.

\bibitem{rev}
F.~J.~Tipler, C.J.S. Clarke, G.F.R. Ellis, ``Singularity and
Horizon - A review article'', in General Relativity and
Gravitation, vol. 2,
(ed. A. Held), Plenum Press, New York, 1980. \\
D.~A.~Konkowski and T.~M.~Helliwell, ``Singularities In Colliding
Gravitational Plane Wave Space-Times,'' Class.\ Quant.\ Grav.\
{\bf 6} (1989) 1847.



\bibitem{prebigbang} A.~Buonanno, T.~Damour and G.~Veneziano,
``Pre-big bang bubbles from the gravitational instability of
generic  string vacua,'' Nucl.\ Phys.\ B {\bf 543} (1999) 275
[arXiv:hep-th/9806230].

\bibitem{fkv}
A.~Feinstein, K.~E.~Kunze and M.~A.~Vazquez-Mozo, ``Initial
conditions and the structure of the singularity in pre-big-bang
cosmology,'' Class.\ Quant.\ Grav.\  {\bf 17} (2000) 3599
[arXiv:hep-th/0002070].

\bibitem{bv}
V.~Bozza and G.~Veneziano, ``O(d,d)-invariant collapse/inflation
from colliding superstring waves,'' JHEP {\bf 0010} (2000) 035
[arXiv:hep-th/0007159].


\bibitem{Gowdy} R.~H.~Gowdy,
``Gravitational Waves In Closed Universes,'' Phys.\ Rev.\ Lett.\
{\bf 27} (1971) 826.

\bibitem{hooft} T.~Dray and G.~'t Hooft,
``The Gravitational Shock Wave Of A Massless Particle,'' Nucl.\
Phys.\ B {\bf 253} (1985) 173.

\bibitem{Ori} A.~Ori and E.~E.~Flanagan,
``How generic are null spacetime singularities?,'' Phys.\ Rev.\ D
{\bf 53} (1996) 1754 [arXiv:gr-qc/9508066].
``Null weak singularities in plane-symmetric spacetimes,'' Phys.\
Rev.\ D {\bf 57} (1998) 4745 [arXiv:gr-qc/9801086].
``Evolution Of Linear Gravitational And Electromagnetic
Perturbations Inside A Kerr Black Hole,'' Phys.\ Rev.\ D {\bf 61}
(2000) 024001.

\bibitem{BS} P.~Bell and P.~Szekeres,
``Interacting Electromagnetic Shock Waves In General Relativity,''
Gen.\ Rel.\ Grav.\  {\bf 5} (1974) 275.

\bibitem{OSc} S.~O'Brien and J.~L.~Synge,
``Jump conditions at discontinuities in General Relativity'',
Comm. Dublin Inst. Adv. Stud. A{\bf 9} (1952).

\bibitem{Gurses1} M.~Gurses and E.~Sermutlu,
``Colliding gravitational plane waves in dilaton gravity,'' Phys.\
Rev.\ D {\bf 52} (1995) 809 [arXiv:hep-th/9503218].

\bibitem{higherBS}
M.~Gurses, Y.~Ipekoglu, A.~Karasu and C.~Senturk, ``Higher
dimensional Bell-Szekeres metric,'' Phys.\ Rev.\ D {\bf 68} (2003)
084007 [arXiv:gr-qc/0305105].

\bibitem{GP} M.~Gutperle and B.~Pioline,
``Type IIB colliding plane waves,'' JHEP {\bf 0309} (2003) 061
[arXiv:hep-th/0308167].

\bibitem{hs}
G.~T.~Horowitz and L.~Susskind,
J.\ Math.\ Phys.\  {\bf 42} (2001) 3152 [arXiv:hep-th/0012037].


\bibitem{pirani} F.~A.~E.~Pirani, ''Introduction to gravitational
raidation theory'', in {\it Lectures on general relativity}, pp.
249-373. Brandeis Summer Institue of Theoretical Physics, {\bf 1},
364. Prentice Hall, Englewood Cliffs.


\end{thebibliography}
